\documentclass[12pt, a4paper]{article}
\usepackage{latexsym,amsfonts,amssymb}
\usepackage{graphicx}
\usepackage{indentfirst}
\usepackage{cite}
\usepackage[cp1251]{inputenc}

\newtheorem{theorem}{Theorem}

\newtheorem{remark}{Remark}

\newcommand{\be} {\begin{equation}}
\newcommand{\ee} {\end{equation}}
\newcommand{\ba} {\begin{array}}
\newcommand{\ea} {\end{array}}

\begin{document}

\begin{center}
{\Large \bf Exact solutions of nonlinear boundary value problems of the Stefan type}

 \medskip

{\bf Roman Cherniha$^{\dag,\ddag}$} {\bf and  Sergii Kovalenko$^\dag$}
 \\
{\it  $^\dag$~Institute of Mathematics, Ukrainian National Academy of Sciences,
\\
3 Tereshchenkivs'ka Street, Kyiv 01601, Ukraine}
\\
{\it $^\ddag$~Lesya Ukrayinka Volyn' National University,
\\
13 Volia Avenue, Lutsk 43025, Ukraine}
\\

\medskip
 E-mail: cherniha@imath.kiev.ua and kovalenko@imath.kiev.ua
\end{center}

\begin{abstract}
The (1+1)-dimensional nonlinear boundary value problem, modeling
the  process of melting and evaporation of metals,
is studied  by means of the classical Lie symmetry method. All
possible Lie operators of the nonlinear heat equation, which allow us
to  reduce the problem to the boundary value problem for the system of
ordinary differential equations, are found. The forms of  heat
conductivity coefficients are established when the given  problem
can be analytically  solved in an explicit form.
\medskip
\end{abstract}

\section{\bf Introduction}

\noindent It is well-known that processes of melting and evaporation of metals in the case where their
surface is exposed to  a powerful flux of energy are described by a
nonlinear boundary value problem of the Stefan type
\cite{ready, alex93, crank84, rub71}. In the (1+1)-dimensional case
the relevant boundary value problem (BVP)  reads as \cite{ch-od90, ch93, ch-od91}:
\begin{eqnarray}
& & \frac{\partial}{\partial
x}\left(\lambda_{1}(T_{1})\frac{\partial T_{1}}{\partial x}\right) =
C_{1}(T_{1}) \frac{\partial
T_{1}}{\partial t}, \ \ \ 0<S_{1}(t)<x<S_{2}(t),\label{1} \\
& & \frac{\partial}{\partial x}\left(\lambda_{2}(T_{2})
\frac{\partial T_{2}}{\partial x}\right) = C_{2}(T_{2})
\frac{\partial
T_{2}}{\partial t}, \ \ \  x>S_{2}(t),\label{2} \\
 & & \qquad x = S_{1}(t):\lambda_{1}(T_{1})
\frac{\partial T_{1}}{\partial x} = \frac{d S_1}{d t} H_{v} - q(t),\label{3} \\
 & & \qquad x  =  S_{1}(t):T_{1} = T_{v},\label{4} \\ & & \qquad x = S_{2}(t):\lambda_{2}(T_{2}) \frac{\partial T_{2}}{\partial x} =
\lambda_{1}(T_{1}) \frac{\partial T_{1}}{\partial x} +
\frac{d S_{2}}{d t} H_{m},\label{5} \\ & & \qquad x = S_{2}(t):T_{1} =
T_{2} = T_{m},\label{6}
\\ & & \qquad x = +\infty: T_{2} = T_{0},\label{7}
\end{eqnarray}
where $T_{v}$, $T_{m}$, $T_{0}$ are the known temperatures of
evaporation, melting, and solid phase of metal, respectively;
$\lambda_{k}$ are thermal conductivities; $C_{k}$, $H_{v}$, $H_{m}$
are specific heat values per unit volume; $q(t)$ is a function
presenting  the energy flux being absorbed by the metal; $S_{k}$ are
the phase division boundary coordinates to be found;
$\frac{d S_{k}}{d t}$ are the phase division
boundary velocities; $T_{k}(t, x)$ are unknown temperature fields;
and index $k = 1, 2$ corresponds to the liquid and solid phases,
respectively.

In this BVP with moving boundaries,  Eqs. (\ref{1}) and (\ref{2})
describe the heat transfer process in liquid and solid phases, respectively, the
boundary conditions (\ref{3}) and (\ref{4}) present evaporation dynamics on the surface $S_{1}$, and
the boundary conditions  (\ref{5}) and (\ref{6}) are the well-known
Stefan conditions on the surface $S_{2}$ dividing the liquid and
solid phases. Assuming that the liquid  phase thickness is
considerably less than the solid phase thickness, one may use the
Dirichlet condition (\ref{7}). It should be stressed that we neglect
the initial distribution of the temperature in the solid phase and
  consider the process at that stage when two phases already take place. This means that we start to
describe the process at time  $t=t^*>0$ when
\begin{eqnarray}
T_{1}=T_l(t), \quad T_{2}=T_s(t),\nonumber
\end{eqnarray}
 where
$T_l(t)$ and $T_s(t)$ are non-constant functions, which are defined
by  the solutions of the  problem  (\ref{1})--(\ref{7}).

The simplest realistic case of this BVP with moving boundaries
occurs under the assumption $q(t)=\mbox{const}$ when the process has a long
quasistationary phase after a short transient phase for $t \in
(0,t^*)$. It means that   the unknown functions $S_{1}$ and $S_{2}$
are  linear with respect to the time if $t>t^*$ and
$S_{2}-S_{1}=\mbox{const}$; therefore, the  BVP  (\ref{1})--(\ref{7}) can
be reduced to the problem for ordinary differential equations by
ansatz
\[
T_{1}=T_1(z), \quad T_{2}=T_2(z), \quad  z=x-vt,
\]
where $v=\frac{d S_{1}}{d t}=\frac{d S_{2}}{d t}>0$ is an unknown phase
division boundary velocity. It turns out that the BVP obtained can
be exactly solved in an implicit form; moreover,  the solution is
expressed in an explicit form for a wide range of functions
$\lambda_{k}$  and $C_{k}$  \cite{ch-od90, ch93}. Note that
the case of constant values $\lambda_{k}$  and $C_{k}$, i.e. the fact that Eqs.
(\ref{1}) and (\ref{2}), are linear heat equations, was considered in
the pioneering paper \cite{du-yar}.

This paper is devoted to finding new reductions of the nonlinear BVP
(\ref{1})--(\ref{7}) to the simpler problems and to constructing
their exact solutions. The main idea is to apply
 the classical Lie symmetry method \cite{ovs, b-k, olv}.
In section 2,  all possible Lie operators of the nonlinear heat
equations (\ref{1}) and (\ref{2}), which allow us to  reduce the problem
to the BVP for an ordinary differential equation system, are found. In
section 3, the forms of  the coefficients arising in BVP (\ref{1})--(\ref{7}) are established when
 the boundary value problems obtained in section 2
 can be analytically  solved in an explicit form and the relevant exact solutions
 are constructed.  Application of the exact  solution obtained  in
 the  case of linear basic equations (see Eqs.
(\ref{1}) and (\ref{2}) with constant coefficients) is presented to
calculate the temperature fields and phase division boundary
coordinates for the parameters, which are typical for aluminium.
Section 4 concludes the paper.

\section{\bf Reduction of the problem to the nonlinear BVP for the system of ODEs}

\noindent It can be noted that BVP (\ref{1})--(\ref{7}) can be simplified if
one applies the Kirchhoff substitution
\begin{equation}\label{8}
u = \int\limits_0^{T_{1}} {C_{1}(T_{1})}\,dT_{1}, \quad v =
\int\limits_0^{T_{2}} {C_{2}(T_{2})}\,dT_{2}.
\end{equation}
Substituting (\ref{8}) into  (\ref{1})--(\ref{7}) and making the
relevant calculations, we arrive at the equivalent BVP of the form
\begin{eqnarray}
& & \frac{\partial u}{\partial t}  =  \frac{\partial}{\partial
x}\left(d_{1}(u) \frac{\partial
u}{\partial x}\right), \ \ \ t>0, \ S_{1}(t)<x<S_{2}(t),\label{9}  \\
 & & \frac{\partial v}{\partial t}  =  \frac{\partial}{\partial
x}\left(d_{2}(v)
\frac{\partial v}{\partial x}\right), \ \ \ t>0, \ x>S_{2}(t),\label{10} \\
& & \qquad x  =  S_{1}(t):d_{1}(u)
\frac{\partial u}{\partial x} = \frac{d S_1}{d t} H_{v} - q(t),\label{11} \\
& & \qquad x  =  S_{1}(t):u = u_{1},\label{12} \\ & & \qquad x  =
S_{2}(t):d_{2}(v) \frac{\partial v}{\partial x} = d_{1}(u)
\frac{\partial u}{\partial x} + \frac{d S_{2}}{d t} H_{m},\label{13}
\\ & & \qquad x  =  S_{2}(t):u = u_{2}, \ v =  v_{2},\label{14}
\\ & & \qquad x  =  +\infty: v = v_{0},\label{15}
\end{eqnarray}
where $d_{1}(u) = \frac{\lambda_{1}(T{1})}{C_{1}(T_{1})}, \ d_{2}(v)
= \frac{\lambda_{2}(T{2})}{C_{2}(T_{2})}.$

Now one can see that BVP (\ref{9})--(\ref{15}) is based on the
standard  nonlinear heat equations (\ref{9}) and (\ref{10}). We want
to find all possible reduction of this BVP to a nonlinear BVP based
on ODEs (not PDEs !) using the known Lie symmetry operators of the
nonlinear heat equation (NHE). We start from theorem 1 that gives
strong restrictions on the form of Lie symmetry operators.

\begin{theorem}
A Lie symmetry operator of NHE
\begin{equation}\label{16}
\frac{\partial u}{\partial t} = \frac{\partial}{\partial
x}\left(d(u) \frac{\partial u}{\partial x}\right), \quad
d(u)\not=const
\end{equation}
reduces this equation together with the moving boundary conditions
\begin{equation}\label{17}
x = S_1(t): u = u_{1},  \qquad x = S_2(t): u = u_{2},
\end{equation}
where $S_k(t)$ are  unknown non-constant functions while $
u_{2}>u_1$ are the given constants, to an ODE with the relevant
boundary condition iff the operator in question  up to local
transformations $x\rightarrow x+x_{0}, \ t\rightarrow t+t_{0} \,
(x_{0}\in \mathbb{R},t_{0} \in \mathbb{R})$ is equivalent either to
\begin{equation}\label{18}
X_{1} = \partial_{t} + \mu \partial_{x}, \ \mu \in \mathbb{R}
\end{equation}
or to
\begin{equation}\label{19}
X_{2} = 2t\partial_{t} + x\partial_{x}.
\end{equation}
\end{theorem}

\noindent \textbf{Proof}. The group classification of NHE (\ref{16})
is well-known \cite{ovs}. If $d(u)$ is an arbitrary function, then
the maximal algebra of invariance (MAI) is generated by the basic
operators  $\langle \partial_{t},
\partial_{x}, 2t \partial_{t} + x
\partial_{x}\rangle$. There are three special cases of extension
of this three-dimensional algebra (see table 1).

\medskip

\textbf{Table 1.} {\bf  Lie algebras of NHE (16).}
{\renewcommand{\arraystretch}{1.5}
\begin{center}
\begin{tabular}{|c|l|l|}
  \hline
  \emph{No.} & \emph{The form of NHE} & \emph{MAI}\\
  \hline\hline
  1. & $u_{t} = (e^u u_{x})_{x}$ & $\langle \partial_{t},
\partial_{x}, 2t \partial_{t} + x
\partial_{x}, x \partial_{x} + 2 \partial_{u} \rangle$  \\
  2. & $u_{t} = (u^k u_{x})_{x}, k\neq 0, -\frac {4}{3}$ & $\langle \partial_{t},
\partial_{x}, 2t \partial_{t} + x
\partial_{x},k x \partial_{x} + 2 u \partial_{u} \rangle$ \\
  3. & $u_{t} = (u^{-\frac {4}{3}} u_{x})_{x}$ & $\langle \partial_{t},
\partial_{x}, 2t \partial_{t} + x
\partial_{x},-\frac {4}{3} x \partial_{x} + 2 u \partial_{u}, x^2 \partial_{x} - 3xu\partial_{u} \rangle$
\\
  \hline
\end{tabular}
\end{center}}

\medskip

Each NHE that admits four- or five-dimensional Lie algebra is
reduced to one of those from table 1 by the equivalence
transformations
\begin{equation}\label{19a}
\bar{t} = e_0t + t_{0},\quad \bar{x} = e_1x + x_{0}, \quad \bar{u} =
e_2u + u_{0},
\end{equation}
where $e_0, t_{0}, e_1, x_{0}, e_2$, and $u_{0}$ are arbitrary group
parameters.

 According to the Lie theory, each linear combination
of the Lie symmetry operators allow us to reduce the relevant NHE to
an ODE; however, we also need the correctly-specified reduction of
the moving boundary condition (\ref{17}).

Let us consider an arbitrary  function $d(u)$ and assume $d(u)\neq
e^{u}, u^k$. In this case, the most general form of the Lie symmetry
operator is
\begin{equation}\label{20}
X_{3} = (\lambda_{1} +
2\lambda_{3}t)\partial_{t}+(\lambda_{2}+\lambda_{3}x)\partial_{x}.
\end{equation}
Hereinafter,  $\lambda$ with indices are arbitrary constants. It is
well-known that  operator  (\ref{20}) up to local transformations of
NHE (\ref{16})
 \begin{equation}\label{20*} x\rightarrow x+x_{0}, \quad t\rightarrow
t+t_{0} \, \left( x_{0}\in \mathbb{R}, t_{0} \in \mathbb{R}\right),
\end{equation}
which is a subset of (\ref{19a}),  can be reduced either to form
 (\ref{18}) (if  $\lambda_{3}=0$) or (\ref{19}) (if  $\lambda_{3}\not=0$).
 Note that the case $\lambda_{3}=0$ and $\lambda_{1}\lambda_{2}=0$ leads
 to the ansatz, which contradicts the moving boundary condition
 (\ref{17}).

Moreover, transformations (\ref{20*}) preserve  the form of
condition (\ref{17}) (up to new notations). Examination  of operator
(\ref{18}) immediately leads to the result of paper \cite{ch93}
because it generates the plane wave ansatz
\begin{equation}
u(t,x)=U(z), \quad  z=x - \mu t,
\end{equation}
which leads only to the  linear form of the function $S(t)$ in
(\ref{17}).

Examination  of operator (\ref{19}) immediately leads to the ansatz
\begin{equation}\label{21}
u=u(\omega), \quad  \omega = \frac{x}{\sqrt{t}}.
\end{equation}
Using the second formula from (\ref{21}), one sees that the moving
boundary conditions (\ref{17}) take the form $\omega =
\frac{S_{k}(t)}{\sqrt{t}}: u = u_{k} \, (k = 1,2)$. Since these
equalities must take place for arbitrary time $t>t^*$, we arrive at
the conditions
\begin{equation}\label{22}
\frac{S_{k}(t)}{\sqrt{t}} = \omega_{k},
\end{equation}
where  $\omega_{k}$ are unknown constants.

\noindent  Thus, ansatz (\ref{21}) reduces  problem  (\ref{16}) and
(\ref{17}) to the problem
\begin{eqnarray}
& & \left(d(u) u_{\omega}\right)_{\omega} + \frac{\omega}{2} u_{\omega} = 0, \\
& & \qquad \omega = \omega_{1}: u = u_{1}, \\ & & \qquad \omega =
\omega_{2}: u = u_{2},
\end{eqnarray}
if  the moving boundary conditions (\ref{17}) take the form
\begin{equation}\label{23}
x = \omega_{k} \sqrt{t}: u = u_{k}.
\end{equation}

If NHE admits four- or five-dimensional Lie algebra, then one can
reduce to the form listed in  case 1, 2 or 3 of table 1 by the
substitution $\bar{u} = e_2u + u_{0}, \, e_2\not=0$ (see
(\ref{19a})). Simultaneously, conditions (\ref{17})
take the form
\[  x = S_1(t): \bar{u} = e_2u_1 + u_{0}=\bar{u}_{1},  \qquad
x = S_2(t): \bar{u} =e_2u_2 + u_{0}= \bar{u}_{2},  \] where
$\bar{u}_{1}-\bar{u}_{2}\not=0$. Thus,  we arrive at the same
problem (\ref{16}) and (\ref{17}) with bars. Hereafter, bars are
omitted so that we need to examine only three cases from table 1.

 Let us
consider the first case of table 1. Here, the most general form of the
Lie symmetry operator is
\begin{equation}\label{26}
X_{4} = (\lambda_{1} +
2\lambda_{3}t)\partial_{t}+(\lambda_{2}+(\lambda_{3}+\lambda_{4})x)\partial_{x}
+ 2\lambda_{4}\partial_{u}.
\end{equation}
Clearly, we should assume $\lambda_{4}\not=0$ otherwise we
arrive at the previous case (see operator  (\ref{20})).

Depending on the values of  $\lambda_{i} \, (i = \overline{1,4}$), the
corresponding system of characteristic equations
\begin{equation}
\frac{dt}{\lambda_{1} + 2\lambda_{3}t} =
\frac{dx}{\lambda_{2}+(\lambda_{3}+\lambda_{4})x} =
\frac{du}{2\lambda_{4}}
\end{equation}
  can generate only two types of ans\"atze
\begin{eqnarray}
& & u = F(\omega) + \Phi(x),\label{27} \\ & & u = F(\omega) +
\Psi(t),\label{28}
\end{eqnarray}
where  $F(\omega)$ is a new unknown function of the known variable
$\omega(t,x)$ and $\Phi(x)$ and  $\Psi(t)$ are the known functions.
Substituting the first  of them into  (\ref{17}) and dealing
similar to the case of operator (\ref{21}), we arrive at the
conclusion that $x = S_{k}(t) \to \omega = \omega_{k}$  so that
(\ref{17}) takes the form
\begin{equation}
\omega = \omega_{k}: F(\omega_{k}) + \Phi(S_{k}(t)) =
u_{k}.\label{29}
\end{equation}
Since these equalities must take place for arbitrary time $t>t^*$, we
arrive at the conditions $\Phi(x)=\Phi(S_{k}(t))=\Phi_k=\mbox{const}$. So
ansatz (\ref{27}) can be rewritten in the form
\begin{equation}\label{30}
u = F^{\ast}(\omega),
\end{equation}
where $F^{\ast}(\omega)=F(\omega) + \Phi_k$. On the other hand,
ansatz (\ref{30}) can be obtained from operator (\ref{26}) only
under condition   $\lambda_{4} = 0$ but we assumed
$\lambda_{4}\not=0$. In the quite similar way, one proves that
application of ansatz (\ref{28}) also leads to the requirement
$\lambda_{4} = 0$. Thus, we have shown that there are no new
reductions of problem  (\ref{16}) and (\ref{17}) in case 1 of
table 1.

In case 2 of table 1, the most general form of the Lie symmetry
operator is
\begin{equation}\label{31}
X_{5} = (\lambda_{1} +
2\lambda_{3}t)\partial_{t}+(\lambda_{2}+(\lambda_{3}+k
\lambda_{4})x)\partial_{x} + 2\lambda_{4} u \partial_{u}.
\end{equation}
Depending on the values of  $\lambda_{i} \, (i = \overline{1,4}$), the
corresponding system of characteristic equations
\begin{equation}
\frac{dt}{\lambda_{1} + 2\lambda_{3}t} =
\frac{dx}{\lambda_{2}+(\lambda_{3}+k \lambda_{4})x} =
\frac{du}{2\lambda_{4} u}
\end{equation}
  can generate only two types of ans\"atze
\begin{eqnarray}
& & u = F(\omega)\cdot \Phi(x), \\ & & u = F(\omega)\cdot \Psi(t).
\end{eqnarray}
Nevertheless, these ans\"atze differ from ans\"atze
(\ref{27}) and (\ref{28}); one can deal with them in a quite similar
way. Finally,  one arrives at the function restrictions
$\Phi(x)=\Phi_0$ and $\Psi(t)=\Psi_0$ (where $\Phi_0 = \mbox{const}$ and
$\Psi_0 = \mbox{const}$), which immediately lead to ansatz (\ref{30}) with
$F^{\ast}(\omega)=F(\omega)\Phi_0$ or $F^{\ast}(\omega)=F(\omega)
\Psi_0$. Thus, there are no new reductions of problem  (\ref{16}) and
(\ref{17}) in case 2 of table 1.

Examination  of case 3 from table 1 is rather cumbersome;
however the result is still the same: problem  (\ref{16}) and
(\ref{17}) with $d(u)= u^{-\frac{4}{3}}$  can be reduced to  an ODE
with the relevant boundary condition iff the Lie symmetry operator has form (\ref{20}).

The proof is now completed. $\blacksquare$

\begin{remark}
To prove the theorem, only the structure of
the corresponding Lie ans\"atze was used. These ans\"atze are listed in an explicit form in \cite{fh-sr-am93}.
\end{remark}

\begin{remark}
One easily checks that operators
(\ref{18}) and (\ref{19}) reduce  problem (\ref{16}) and (\ref{17}) to
the linear ODE with the relevant boundary condition also in the case
$d(u)=const$. However, this is rather a long routine to prove that
there are no new Lie symmetry operators providing the same
reductions because the linear heat equation admits
infinity-dimensional Lie algebra.
\end{remark}

\begin{theorem}
A Lie symmetry operator of NHE reduces the nonlinear BVP (\ref{9})--(\ref{15}) to a BVP for two ODEs  with the relevant boundary
conditions iff the operator in question up to local transformations
$x\rightarrow x+x_{0}, \ t\rightarrow t+t_{0} \, (x_{0}\in
\mathbb{R},t_{0} \in \mathbb{R})$ is equivalent either to operator
(\ref{18}) or
 to (\ref{19}) and the functions $S_k(t), \, k=1,2, $ and $q(t)$
  have the correctly specified forms
  \begin{equation}\label{45}
S_1= \mu t +\omega_1, \ S_2= \mu t +\omega_2, \ q(t)=q_0
\end{equation}
or
 \begin{equation}\label{44}
S_{1} = \omega_{1} \sqrt{t}, \quad S_{2} = \omega_{2} \sqrt{t},
\quad q(t) = \frac{q_{0}}{\sqrt{t}},
\end{equation}
where $\mu$ and $\omega_{k}\, k=1,2$  are to-be-determined constants
and $q_0$ is an arbitrary positive constant.
\end{theorem}

\noindent \textbf{Proof.} The proof of this theorem is based on
theorem 1. One notes that the nonlinear BVP (\ref{9})--(\ref{15})
contains NHE (\ref{9}) with the boundary conditions (\ref{12}),
(\ref{14}) and NHE (\ref{10}) with the boundary conditions (\ref{14}),
(\ref{15}) so that this BVP can be reduced to a BVP for ODEs only
in the case when the given Lie symmetry operator up to local
transformations $x\rightarrow x+x_{0}, t\rightarrow t+t_{0},
x_{0}\in \mathbb{R}, t_{0} \in \mathbb{R}$ is equivalent either to
operator (\ref{18}) or to (\ref{19}).

To complete the proof, we need to check whether these operators
correctly reduce the boundary conditions (\ref{11}) and  (\ref{13}). In paper \cite{ch93}, this has been shown
for operator (\ref{18}) and it was established that the phase division
lines $S_k(t)= \mu t +\omega_k, \, k=1,2, $ where the constants
$\mu $ and $ \omega_k$ are to-be-determined.

The application of operator (\ref{19}) to BVP (\ref{9})--(\ref{15})
leads to the ansatz
\begin{equation}\label{21*}
u=u(\omega), \quad v=v(\omega), \quad  \omega = \frac{x}{\sqrt{t}}.
\end{equation}
To satisfy the boundary conditions (\ref{12}) and (\ref{14}), we
obtain  the phase division
 lines of form (\ref{22}), i.e.
 \begin{equation}\label{22*}
\frac{S_1(t)}{\sqrt{t}} = \omega_{1}, \quad
\frac{S_2(t)}{\sqrt{t}}=\omega_{2},
\end{equation}
where  $\omega_{k}, \, k=1,2$  are to-be-determined constants. The
direct calculations show that ansatz (\ref{21*}) correctly reduces
the boundary conditions (\ref{11}) and  (\ref{13}) with the
restriction (\ref{22*}) if additionally the energy flux is given by
the function
\begin{equation}\label{46}
q(t) = \frac{q_{0}}{\sqrt{t}}.
\end{equation}

Thus, substituting formulae (\ref{21*})--(\ref{46})
into BVP  (\ref{9})--(\ref{15}) and making the relevant
simplifications, we arrive at the BVP for two ODEs of the form
\begin{eqnarray}
& & (d_{1}(u) u_{\omega})_{\omega} + \frac{\omega}{2} u_{\omega} = 0,\label{36} \\
& & (d_{2}(v) v_{\omega})_{\omega} + \frac{\omega}{2} v_{\omega} = 0,\label{37} \\
&& \qquad \omega = \omega_{1}: d_{1}(u) u_{\omega} =
\frac{\omega_{1}}{2}
H_{v} - q_{0},\label{38} \\ & & \qquad \omega = \omega_{1}: u=u_{1},\label{39} \\
& & \qquad \omega = \omega_{2}: d_{2}(v) v_{\omega} = d_{1}(u)
u_{\omega} + \frac{\omega_{2}}{2} H_{m},\label{40} \\ & & \qquad
\omega = \omega_{2}: u = u_{2}, \ v = v_{2},\label{41} \\ & & \qquad
\omega = + \infty: v = v_{0}.\label{42}
\end{eqnarray}
 The proof is now completed. $\blacksquare$

\section{\bf Exact solutions of the nonlinear BVP (\ref{1})--(\ref{7})}

\noindent It was established in the previous section that ansatz (\ref{21*})
reduces BVP (\ref{1})--(\ref{7})  to the BVP for two ODEs
(\ref{36})--(\ref{42})  under the corresponding restrictions.
Nevertheless, BVP (\ref{36})--(\ref{42}) is much simpler than the
original problem, this BVP cannot be exactly  solved in the general
case because nonlinear ODEs (\ref{36}) and  (\ref{37}) are
integrable only in special cases. Here, we consider such cases in
details.

First of the all, we introduce new variables using the well-known
formulae
\begin{equation}\label{48}
U = \int\limits_{u_{*}}^u {d_{1}(u)}\, du, \quad V =
\int\limits_{v_{*}}^v {d_{2}(v)}\, dv,
\end{equation}
(we assume that $U$ and $V$ are continuous on $[u_{*}, +\infty)$ and
$[v_{*}, +\infty)$, respectively).

\noindent The local substitution (\ref{48}) reduces BVP (\ref{36})--(\ref{42}) to the form
\begin{eqnarray}
& & U_{{\omega}{\omega}} + \frac{\omega}{2} D_{1}(U) U_{\omega} = 0,\label{49} \\
& & V_{{\omega}{\omega}} + \frac{\omega}{2} D_{2}(V) V_{\omega} = 0,\label{50} \\
& & \qquad \omega = \omega_{1}: U_{\omega} = \frac{\omega_{1}}{2}
H_{v} - q_{0},\label{51} \\ & & \qquad \omega = \omega_{1}: U=U_{1},\label{52} \\
& & \qquad \omega = \omega_{2}: V_{\omega} = U_{\omega} +
\frac{\omega_{2}}{2} H_{m},\label{53} \\ & & \qquad \omega =
\omega_{2}: U = U_{2}, \ V = V_{2},\label{54} \\ & & \qquad \omega =
+ \infty: V = V_{0},\label{55}
\end{eqnarray}
where the functions $U(\omega), V(\omega)$ and constants
$\omega_{1}, \omega_{2}$ are found and  $D_{1}(U) =
\frac{1}{d_{1}(u)}$, $D_{2}(V) = \frac{1}{d_{2}(v)}$; \ $U_{k} =
\int\limits_{u{*}}^{u_{k}} {d_{1}(u)}\, du$,  $V_{k} =
\int\limits_{v{*}}^{v_{k}} {d_{2}(v)}\, dv$, $k=0,1,2.$

Since the basic  Eqs. (\ref{49}) and  (\ref{50}) are still nonlinear
second-order ODEs, we used the book \cite{pol-za}, which is the
essential extension of the classical Kamke handbook, to specify the
integrable cases. Taking into account that $\omega, D_{1}(U)$ and
$D_{2}(V)$ should be positive (otherwise one obtains non-realistic
equations for the given process), only three cases were separated,
which are listed in table 2. Now one notes that the general
solutions presented in table 2 can be applied for analytically solving
BVP (\ref{49})--(\ref{55}) in nine different cases. Let us consider
the most typical of them.

\newpage

\textbf{Table 2.} {\bf Solutions of ODEs of the form  (53).}
{\renewcommand{\arraystretch}{1.6}
\begin{center}
\begin{tabular}{|c|l|l|}
  \hline
  \emph{ No } & \emph{ODE} & \emph{General solution}\\
  \hline\hline
  1. & $U_{{\omega}{\omega}} + \frac{a^2}{2} \omega U_{\omega} = 0$ $(a = const)$ & $U = C_{2} + C_{1} \frac{\sqrt{\pi}}{a} \textnormal{erf} \left(\frac{a\omega}{2}\right)$  \\
  2. & $U_{{\omega}{\omega}} + \frac{a^2}{2} \frac{1}{U^2}\omega U_{\omega} = 0$ $(a = const)$ & $U = C_{1}\left(\frac{\sqrt{\pi}}{2}
   \textnormal{erf}(\tau) + C_{2}\right), \omega = \frac{1}{a} \left(2\tau U + C_{1} e^{-{\tau}^2}\right)$ \\
  3. & $U_{{\omega}{\omega}} + \frac{a^2}{2} e^U \omega U_{\omega} = 0$ $(a = const)$ & $U = \int \frac{d\tau}{f(\tau)} + C_{2}, \omega = \tau
  e^{-\frac{U}{2}}$, \\ & & $C_{1} = \ln (2f - \tau) - \frac{\tau}{2f - \tau} - \frac{a^2}{4}{\tau}^2$ \\
  \hline
\end{tabular}
\end{center}}

\medskip

\noindent \textbf{Example 1:} $D_{1}(U) = a^2, \  D_{2}(V) = b^2$
(hereafter $a$ and $ b$  are arbitrary positive  constants).

\medskip

\noindent According to case 1 of table 2  the
general solutions of  Eqs. (\ref{49}) and  (\ref{50}) are given by
the formulae
\begin{eqnarray}
& & U = C_{2} + C_{1} \frac{\sqrt{\pi}}{a} \, \textnormal{erf}
\left(\frac{a\omega}{2}\right),\label{60} \\ & & V = C_{4} + C_{3}
\frac{\sqrt{\pi}}{b} \, \textnormal{erf}
\left(\frac{b\omega}{2}\right).\label{61}
\end{eqnarray}
Hereafter, $C_{i} (i=1,\dots,4)$ are to-be-determined  constants.

Substituting solution   (\ref{60}) into the boundary conditions
(\ref{52}) and (\ref{54}), one finds the constants $C_{1}$ and
$C_{2}$:
\begin{equation}\label{62}
C_{1} = \frac{a}{\sqrt{\pi}}\frac{U_{2} -
U_{1}}{\textnormal{erf}\left(\frac{a\omega_{2}}{2}\right) -
\textnormal{erf}\left(\frac{a\omega_{1}}{2}\right)}, \quad  C_{2} =
\frac{U_{1}\,\textnormal{erf}\left(\frac{a\omega_{2}}{2}\right) -
U_{2}\,\textnormal{erf}\left(\frac{a\omega_{1}}{2}\right)}{\textnormal{erf}\left(\frac{a\omega_{2}}{2}\right)
-\textnormal{erf}\left(\frac{a\omega_{1}}{2}\right)}. \end{equation}
Similarly, the constants $C_{3}$ and $C_{4}$ are found using
formulae (\ref{61}),(\ref{54}) and (\ref{55}):
\begin{equation}\label{63}
C_{3} =\frac{b}{\sqrt{\pi}}\frac{V_{2} -
V_{0}}{\textnormal{erf}\left(\frac{b\omega_{2}}{2}\right) - 1},
\quad  C_{4} =
\frac{V_{0}\, \textnormal{erf}\left(\frac{b\omega_{2}}{2}\right) -
V_{2}}{\textnormal{erf}\left(\frac{b\omega_{2}}{2}\right) - 1}.
\end{equation}
So, substituting formulae (\ref{62}) and  (\ref{63}) into
(\ref{60}) and  (\ref{61}), respectively, we find the unknown
functions in the explicit form
\begin{eqnarray}
& & U =
\frac{U_{1}\,\textnormal{erf}\left(\frac{a\omega_{2}}{2}\right) -
U_{2}\,\textnormal{erf}\left(\frac{a\omega_{1}}{2}\right) + (U_{2} -
U_{1})\,\textnormal{erf}\left(\frac{a\omega}{2}\right)}{\textnormal{erf}\left(\frac{a\omega_{2}}{2}\right)
- \textnormal{erf}\left(\frac{a\omega_{1}}{2}\right)},\label{56}
\\ & & V =\frac{V_{0}\,\textnormal{erf}\left(\frac{b\omega_{2}}{2}\right) -
V_{2} + (V_{2} -
V_{0})\,\textnormal{erf}\left(\frac{b\omega}{2}\right)}{\textnormal{erf}\left(\frac{b\omega_{2}}{2}\right)
- 1}.\label{57}
\end{eqnarray}
However, we also need to specify the parameters $\omega_{1}$ and
$\omega_{2}$, which allow us to find the moving boundaries. This can be
done by substituting  (\ref{56}) and (\ref{57}) into the boundary
conditions (\ref{51})and  (\ref{53}) and taking into account the
equality  $\frac{d}{dx}(\textnormal{erf}(x)) = \frac{2}{\sqrt{\pi}}\,
e^{-x^2}$. After the corresponding calculations, we arrive at the
transcendent equation system
\begin{eqnarray}
& & \frac{a}{\sqrt{\pi}}\frac{U_{2} -
U_{1}}{\textnormal{erf}\left(\frac{a\omega_{2}}{2}\right) -
\textnormal{erf}\left(\frac{a\omega_{1}}{2}\right)}\,
e^{-\frac{a^2\omega_{1}^2}{4}} = \frac{\omega_{1}}{2} H_{v} -
q_{0},\label{65}
\\ & & \frac{b}{\sqrt{\pi}}\frac{V_{2} -
V_{0}}{\textnormal{erf}\left(\frac{b\omega_{2}}{2}\right) - 1}\,
e^{-\frac{b^2\omega_{2}^2}{4}} = \frac{a}{\sqrt{\pi}}\frac{U_{2} -
U_{1}}{\textnormal{erf}\left(\frac{a\omega_{2}}{2}\right) -
\textnormal{erf}\left(\frac{a\omega_{1}}{2}\right)}\,
e^{-\frac{a^2\omega_{2}^2}{4}} + \frac{\omega_{2}}{2}
H_{m}\label{65*}
\end{eqnarray}
to find the parameters $\omega_{1}$ and $\omega_{2}$. Thus, formulae
(\ref{56}) and (\ref{57}) present the exact solution of BVP
(\ref{49})--(\ref{55}) with $D_{1}(U) = a^2, D_{2}(V) = b^2.$

Here, we present the application of formulae (\ref{56}) and
(\ref{57}) for solving this BVP with the coefficients, which are
typical for aluminium \cite{ch-od91}. The system (\ref{65}) and
(\ref{65*}) was solved by means of the program MATHEMATICA 5.2:
$\omega_{1}\approx 0.0127, \omega_{2}\approx 0.0202$. With  the
values $\omega_{1}$ and $\omega_{2}$,  the temperature fields for
liquid and solid phases were plotted using the program MAPLE 12 (see
Fig.1).

\begin{figure}[h]
\begin{center}
\begin{minipage}[h]{9cm}
\centerline{\includegraphics[width=9cm]{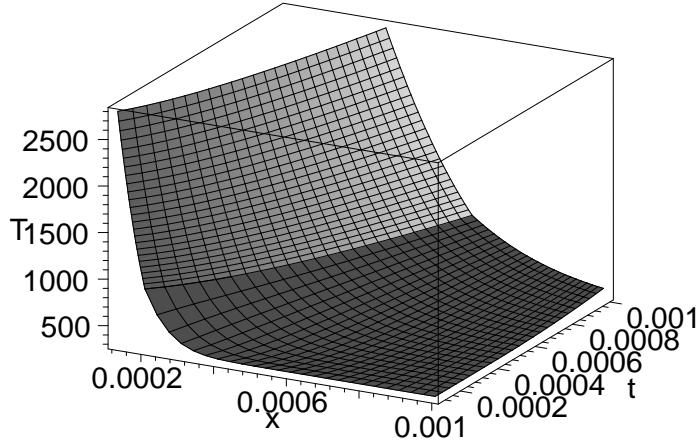}}
\end{minipage}
\end{center}
\center\caption{ Exact solution of the problem (\ref{1})--(\ref{7})
with parameters that are typical for aluminium: $\lambda_{1} =
\lambda_{2} = 240\, \frac{W}{K\cdot m}, C_{1} = 2,7\cdot10^{6}\,
\frac{J}{K\cdot m^{3}}, C_{2} = 2,74\cdot10^{6}\, \frac{J}{K\cdot
m^{3}}, H_{v} = 2,69\cdot10^{10}\, \frac{J}{m^{3}}, H_{m} =
0,17\cdot10^{10}\, \frac{J}{m^{3}}, T_{0} = 300\, K, T_{m} = 933\,
K$ and $T_{v} = 2793\, K$ ; the energy flux was set $q
=\frac{2,5\cdot10^{8}}{\sqrt t}\, \frac{W}{m^{2}}.$}
\end{figure}

\noindent \textbf{Example 2:} $D_{1}(U) = \frac{a^2}{U^2}, \
D_{2}(V) = \frac{b^2}{V^2}$.

\medskip

\noindent According to case 2 of table 2,  the general solutions of  Eqs.
(\ref{49}) and  (\ref{50}) are
\begin{eqnarray}
& & U = C_{1}\left(\frac{\sqrt{\pi}}{2}\, \textnormal{erf}(\tau) +
C_{2}\right), \quad \omega = \frac{1}{a} \left(2\tau U + C_{1}
e^{-{\tau}^2}\right),\label{69} \\ & & V =
C_{3}\left(\frac{\sqrt{\pi}}{2}\, \textnormal{erf}(\nu) +
C_{4}\right), \quad  \omega = \frac{1}{b} \left(2\nu V + C_{3}
e^{-{\nu}^2}\right).\label{70}
\end{eqnarray}
This is important to note that the second formula in (\ref{69})
gives one-to-one correspondence between $\omega$ and  $\tau$,
because the differentiable  function $\omega(\tau)$ is strictly
monotonic:
\begin{equation}
\frac{d \omega}{d \tau} = \frac{1}{a}\left(2 U + 2 \tau \frac{d U}{d
\tau} + C_{1}(-2 \tau)e^{-{\tau}^2}\right) = \frac{2}{a}\,U > 0.
\end{equation}
Hence, the function  $\omega(\tau)$ is reversible and an
inverse strictly monotonic function   $\tau = \tau(\omega)$ exists for all
$\omega > 0$. Analogously, we prove the existence of an inverse
strictly monotonic function   $\nu = \nu(\omega)$ for the function
$\omega(\nu)$ arising in the second formula of (\ref{70}). With
the monotonic differentiable functions   $\tau(\omega)$ and
$\nu(\omega)$, we transform the boundary conditions (\ref{51})--(\ref{55}) with $D_{1}(U) = \frac{a^2}{U^2}, D_{2}(V) =
\frac{b^2}{V^2}$ to the form
\begin{eqnarray}
& & \tau = \tau_{1}:
\frac{dU}{d\tau}\left(\frac{d\omega}{d\tau}\right)^{-1} =
\frac{\omega_{1}}{2}
H_{v} - q_{0},\label{72} \\ & & \tau = \tau_{1}: U=U_{1},\label{73} \\
& & \tau = \tau_{2}, \nu = \nu_{2}:
\frac{dV}{d\nu}\left(\frac{d\omega}{d\nu}\right)^{-1} =
\frac{dU}{d\tau}\left(\frac{d\omega}{d\tau}\right)^{-1} +
\frac{\omega_{2}}{2} H_{m},\label{74} \\ & & \tau = \tau_{2}, \nu = \nu_{2}: U = U_{2}, \  V = V_{2},\label{75} \\
& & \nu = + \infty: V = V_{0}\label{76}.
\end{eqnarray}
Substituting solutions (\ref{69}) and  (\ref{70})   into the
boundary conditions (\ref{73}), (\ref{75}) and  (\ref{76}), one
finds the constants  $C_{i} (i = 1,...,4)$:
\begin{eqnarray}
& & C_{1} = \frac{2}{\sqrt{\pi}}\frac{U_{2} -
U_{1}}{\textnormal{erf}(\tau_{2}) - \textnormal{erf}(\tau_{1})}, \
C_{2} = \frac{\sqrt{\pi}}{2} \frac{U_{1}\,\textnormal{erf}(\tau_{2}) -
U_{2}\,\textnormal{erf}(\tau_{1})}{U_{2} - U_{1}},\label{77}
\\ & & C_{3} =\frac{2}{\sqrt{\pi}}\frac{V_{2} -
V_{0}}{\textnormal{erf}(\nu_{2}) - 1}, \ C_{4} =
\frac{\sqrt{\pi}}{2} \frac{V_{0}\,\textnormal{erf}(\nu_{2}) -
V_{2}}{V_{2} - V_{0}}.\label{78}
\end{eqnarray}
So, substituting formulae (\ref{77}) and  (\ref{78}) into
(\ref{69}) and (\ref{70}), respectively, we find the unknown
functions in the explicit form
\begin{eqnarray}
& & U = U_{1} + \frac{\textnormal{erf}(\tau) -
\textnormal{erf}(\tau_{1})}{\textnormal{erf}(\tau_{2}) -
\textnormal{erf}(\tau_{1})} \,(U_{2} - U_{1}), \ \ \omega =
\frac{2}{a}\left(\tau U + \frac{1}{\sqrt{\pi}}\frac{U_{2} -
U_{1}}{\textnormal{erf}(\tau_{2}) - \textnormal{erf}(\tau_{1})}\,
e^{-\tau^2}\right),\label{79} \\ & & V = V_{2} +
\frac{\textnormal{erf}(\nu) -
\textnormal{erf}(\nu_{2})}{\textnormal{erf}(\nu_{2}) - 1} \,(V_{2} -
V_{0}), \ \ \omega = \frac{2}{b}\left(\nu V +
\frac{1}{\sqrt{\pi}}\frac{V_{2} - V_{0}}{\textnormal{erf}(\nu_{2}) -
1} \,e^{-\nu^2}\right),\label{80}
\end{eqnarray}
where   $\tau_{1}, \tau_{2}, \nu_{2}$ are to-be-determined
parameters. To find these positive parameters, one needs to
substitute the functions $U$ and $V$   into  (\ref{72}) and
(\ref{74}), respectively. Moreover, the condition (see formulae
(\ref{54}))
\begin{equation}
\omega(\tau_{2}) = \omega(\nu_{2}) = \omega_{2}
\end{equation}
should be taken into account. Finally, we arrive at  the
transcendent equation system
\begin{eqnarray}
& & \frac{U_{2} - U_{1}}{\sqrt{\pi}}\left(\frac{a^2}{U_{1}} -
H_{v}\right)\frac{e^{-{\tau}_{1}^2}}{\textnormal{erf}(\tau_{2}) -
\textnormal{erf}(\tau_{1})} = U_{1} H_{v}\tau_{1} - a
q_{0},\nonumber
\\ & & \frac{ab}{\sqrt{\pi}}\frac{V_{2} -
V_{0}}{V_{2}}\frac{e^{-{\nu}_{2}^2}}{\textnormal{erf}(\nu_{2}) - 1}
= \frac{U_{2} - U_{1}}{\sqrt{\pi}}\left(\frac{a^2}{U_{2}} +
H_{m}\right)\frac{e^{-\tau_{2}^2}}{\textnormal{erf}(\tau_{2}) -
\textnormal{erf}(\tau_{1})} + U_{2}H_{m}\tau_{2},\label{67} \\
& & \frac{1}{a}\left(\tau_{2} U_{2} +
\frac{1}{\sqrt{\pi}}\frac{U_{2} - U_{1}}{\textnormal{erf}(\tau_{2})
- \textnormal{erf}(\tau_{1})} \,e^{-{\tau}_{2}^2}\right) =
\frac{1}{b}\left(\nu_{2} V_{2} + \frac{1}{\sqrt{\pi}}\frac{V_{2} -
V_{0}}{\textnormal{erf}(\nu_{2}) - 1}\,
e^{-{\nu}_{2}^2}\right)\nonumber
\end{eqnarray}
for finding  the parameters $\tau_{1}, \tau_{2}$ and $ \nu_{2}.$

Thus, BVP (\ref{49})--(\ref{55}) with $D_{1}(U) = \frac{a^2}{U^2},
D_{2}(V) = \frac{b^2}{V^2}$ has the exact solution (\ref{79}),
(\ref{80}) and
\begin{eqnarray}\label{68*}
& & \omega_1=\frac{2}{a}\left(\tau_{1} U_{1} +
\frac{1}{\sqrt{\pi}}\frac{U_{2} - U_{1}}{\textnormal{erf}(\tau_{2})
- \textnormal{erf}(\tau_{1})} \,e^{-{\tau}_{1}^2}\right),\\
& & \omega_2=\frac{2}{a}\left(\tau_{2} U_{2} +
\frac{1}{\sqrt{\pi}}\frac{U_{2} - U_{1}}{\textnormal{erf}(\tau_{2})
- \textnormal{erf}(\tau_{1})} \,e^{-{\tau}_{2}^2}\right),
\end{eqnarray}
 where the
parameters $\tau_{1}, \tau_{2}$ and $ \nu_{2}$ are found from system (\ref{67}).

\medskip \medskip

\noindent \textbf{Example 3:} $D_{1}(U) = a^2, D_{2}(V) = b^2 e^V$,
i.e., we consider the case when the basic equations (\ref{49}) and
(\ref{50}) are essentially different.

\medskip

\noindent According to cases 1 and  3 of table 2,  the general solutions
of Eqs. (\ref{49}) and  (\ref{50}) are given by the formulae
\begin{eqnarray}
& & U = C_{2} + C_{1} \frac{\sqrt{\pi}}{a} \, \textnormal{erf}
\left(\frac{a\omega}{2}\right),\label{85} \\ & & V = C_{4} +
\int\limits_c^{\nu} \frac{d\nu}{g(\nu)}, \ \omega = \nu
e^{-\frac{V}{2}}, \ C_{3} = \ln (2g - \nu) - \frac{\nu}{2g - \nu} -
\frac{b^2}{4}{\nu}^2,\label{86}
\end{eqnarray}
where  $c $ is an arbitrary  constant.

In a quite similar way as was done in Example 2, one shows that
the function $\omega(\nu)$ is reversible and an inverse
strictly monotone function  $\nu = \nu(\omega)$ exists for all $\omega >
0$. So, using the monotonic differentiable function $\nu(\omega)$,
we transform the boundary conditions (\ref{51})--(\ref{55}) with
$D_{1}(U) = a^2, D_{2}(V) = b^2 e^V $ to the form
\begin{eqnarray}
& & \omega = \omega_{1}: U_{\omega} = \frac{\omega_{1}}{2}
H_{v} - q_{0},\label{87} \\ & & \omega = \omega_{1}: U=U_{1},\label{88} \\
& & \nu = \nu_{2}:
\frac{dV}{d\nu}\left(\frac{d\omega}{d\nu}\right)^{-1} = U_{\omega} +
\frac{\omega_{2}}{2} H_{m},\label{89} \\ & & \nu = \nu_{2}: U =
U_{2}, \ V = V_{2},\label{90} \\ & & \nu = + \infty: V =
V_{0},\label{91}
\end{eqnarray}
where $\omega_{2} = \nu_{2} e^{-\frac{V_{2}}{2}}$.

 The constants  $C_{1}, C_{2}$ and $C_{4}$ are defined by substitution
 (\ref{85}) and (\ref{86}) into the boundary conditions (\ref{88}) and
 (\ref{90}). So, we obtain after the corresponding calculations
\begin{eqnarray}
& & C_{1} = \frac{a}{\sqrt{\pi}}\frac{U_{2} -
U_{1}}{\textnormal{erf}\left(\frac{a\omega_{2}}{2}\right) -
\textnormal{erf}\left(\frac{a\omega_{1}}{2}\right)},\nonumber \\
 & & C_{2} =
\frac{U_{1}\,\textnormal{erf}\left(\frac{a\omega_{2}}{2}\right) -
U_{2}\,\textnormal{erf}\left(\frac{a\omega_{1}}{2}\right)}{\textnormal{erf}\left(\frac{a\omega_{2}}{2}\right)
-\textnormal{erf}\left(\frac{a\omega_{1}}{2}\right)},\label{93} \\
& & C_{4} = V_{2} - \int\limits_c^{\nu_{2}}
\frac{d\nu}{g(\nu)}.\nonumber
\end{eqnarray}
The constant  $C_{3}$ can be  found  using the third equation of
(\ref{86}) with $\nu = \nu_{2}$:
\begin{equation}\label{95}
C_{3} = \ln (2g(\nu_{2}) - \nu_{2}) - \frac{\nu_{2}}{2g(\nu_{2}) -
\nu_{2}} -\frac{b^2}{4}\nu_{2}^2.
\end{equation}
Substituting  (\ref{93}) into  (\ref{85}) and (\ref{86}), we obtain
\begin{eqnarray}
& & U =
\frac{U_{1}\,\textnormal{erf}\left(\frac{a\omega_{2}}{2}\right) -
U_{2}\,\textnormal{erf}\left(\frac{a\omega_{1}}{2}\right) + (U_{2} -
U_{1})\,\textnormal{erf}\left(\frac{a\omega}{2}\right)}{\textnormal{erf}\left(\frac{a\omega_{2}}{2}\right)
- \textnormal{erf}\left(\frac{a\omega_{1}}{2}\right)},\label{96} \\
& & V = V_{2} + \int\limits_{\nu_{2}}^{\nu} \frac{d\nu}{g(\nu)},\ \
\omega = \nu e^{-\frac{V}{2}}.\label{97}
\end{eqnarray}
Finally, substituting (\ref{96}), (\ref{97}) and (\ref{95}) into the
boundary conditions (\ref{87}), (\ref{89}) and (\ref{91}) and making
the corresponding simplification, we arrive at the transcendent
equation system
\begin{eqnarray}
& & \frac{a}{\sqrt{\pi}}\frac{U_{2} -
U_{1}}{\textnormal{erf}\left(\frac{a\omega_{2}}{2}\right) -
\textnormal{erf}\left(\frac{a\omega_{1}}{2}\right)}\,
e^{-\frac{a^2\omega_{1}^2}{4}} = \frac{\omega_{1}}{2} H_{v} -
q_{0},\nonumber
\\ & & V_{0} = V_{2} + \int\limits_{\nu_{2}}^{\infty}
\frac{d\nu}{g(\nu)}\nonumber, \\ & & \ln(2g(\nu) - \nu) -
\frac{\nu}{2g(\nu) - \nu} - \frac{b^2}{4}\nu^2 = \ln(2g_{2} -
\nu_{2}) - \frac{\nu_{2}}{2g_{2} - \nu_{2}} -
\frac{b^2}{4}{\nu_{2}}^2,\nonumber
\end{eqnarray}
where $g_{2} \equiv g(\nu_{2}) = \frac{\nu_{2}}{2} +
e^{\frac{V_{2}}{2}}\left(\frac{\omega_{2}}{2} H_{m} +
\left(\frac{\omega_{1}}{2} H_{v} - q_{0}\right)\,
e^{\frac{a^2}{4}(\omega_{1}^2 - \omega_{2}^2)}\right)^{-1}$.

\section{\bf Conclusions }

\noindent In this paper, the (1+1)--dimensional nonlinear boundary value
problem (\ref{9})--(\ref{15}), modeling the  process of melting and
evaporation of metals,
is studied  by means of the classical Lie symmetry method.  Theorem
2 that gives all possible Lie operators, which allow us to  reduce the
problem to the BVP for the ODE system, was proved.  The forms of  heat
conductivity coefficients are established when the given problem can
be analytically  solved in an explicit form and the relevant exact
solutions are constructed (see the formulae in examples 1--3).

 We found  that the case of a free  boundary, which moves proportionally
to $\sqrt t$, was earlier established for the classical Stefan
problem  with one moving boundary (solidification process)
\cite{tr81, br-tr02}. In the particular case, one notes that
formulae (\ref{56})--(\ref{65*}) with $\omega_1=0$ produce the
corresponding solution obtained in \cite{tr81} for the Stefan
problem  with one moving boundary.  Similarly,  formulae (\ref{79}),
(\ref{80}) and (\ref{68*})  are generalizations of those from
\cite{br-tr02}(see theorem 6) to the case of BVP with two moving
boundaries. It should be noted that the authors of
\cite{tr81, br-tr02} did not use any Lie symmetries but an
assumption that $q(t) = \frac{q_{0}}{\sqrt{t}}$. In paper
\cite{voller04}, a one-phase Stefan  problem based on the linear heat
equation was analytically solved using the above-mentioned
assumption on the free boundary describing the movement of the
shoreline. In \cite{alex93},  exact solutions are presented for
several Stefan problems with different types of boundaries but with
the linear basic equations.

To the best of our knowledge, there are only a few  papers
 devoted to constructions of exact solutions  of nonlinear boundary
value problems of the Stefan type by means of the Lie symmetry method.
Probably, paper \cite{bl-1974} can be quoted as  the first in this
direction. Nevertheless, a Stefan type problem seems to be  a more
complicated object than the standard BVP with the fixed boundaries;
one can note that the Lie symmetry method should be more applicable
just for solving problems with moving boundaries. In fact, the
structure of such boundaries may depend on invariant variable(s) and
this gives a possibility of reducing the given BVP to one of lower
dimensionality. The work is in progress to construct exact solutions
of a {\it a multidimensional BVP} using this approach.


\begin{thebibliography}{99}


\bibitem {ready} Ready J F 1971 {\it Effects of High-power Laser
Radiation} (New York: Academic Press)


\bibitem {rub71} Rubinstein L I 1971 {\it The Stefan problem} (American Mathematical Society:
Providence)


\bibitem {crank84} Crank J 1984 {\it Free and moving boundary problems} (Clarendon Press: Oxford)

\bibitem {alex93} Alexiades V and Solomon A 1993 {\it Mathematical
modeling of melting and freezing processes} (Washington: Hemisphere
publishing corporation)

\bibitem {ch-od90} Cherniha R M  and Odnorozhenko I G 1990 Exact  solutions  of a nonlinear boundary value problem of melting and evaporation of metals under the action of high energy flux
{\it Dopovidi Akad. Nauk Ukrainy (Reports of Acad. Sci. of Ukraine),
ser.A} \textbf{12} 44--47 (in Ukrainian, Summary in English)

\bibitem {ch-od91}  Cherniga R M  and Odnorozhenko I G 1991 Studies of
 the processes of melting and evaporation of metals under the action
  of laser radiation pulses {\it Promyshlennaya Teplotekhnika (Industrial Heat Technic)} \textbf{13} 51--59 (in Russian, Summary in English)

\bibitem {ch93}  Cherniha  R M and Cherniha  N D 1993 Exact  solutions  of
  a class of nonlinear boundary value problems with moving boundaries {\it J. Phys. A.: Math. Gen.} \textbf{26} L935--L940

\bibitem {du-yar} Dulnyov G N and Yaryshev N A 1967 Estimation of process of heat-mass transfer by interaction of the energy flux with a substance
{\it Teplofizika vysokikh temperatur (Heatphysics of High
Temperature)} \textbf{5} 322--328 (in Russian)

\bibitem{ovs}  Ovsiannikov L V 1980 {\it The Group Analysis of
Differential Equations}  (New York: Academic Press)

\bibitem {olv}  Olver P 1986 {\it Applications of Lie Groups to Differential
Equations} (Berlin: Springer)

\bibitem{b-k}   Bluman G W and Kumei S  1989 {\it Symmetries and Differential
 Equations} (Berlin: Springer)


\bibitem {fh-sr-am93} Fushchych W I, Serov M I and Amerov T K 1993 Nonlocal ansatze and solutions of nonlinear system of heat equatins
{\it Ukrainian Math. J.} \textbf{45} 293--302

\bibitem {pol-za} Polyanin A D and Zaitsev V F 2003 {\it Handbook of
exact solutions for ordinary differential equations} (Boca Raton,
FL: CRC Press Company)

\bibitem {tr81} Tarzia D A 1981 An inequality for the coefficient $\sigma$ of the free boundary $s(t) = 2 \sigma \sqrt{t}$ of the Neumann solution for the two-phase Stefan problem {\it Quart. Appl. Math.} \textbf{39} 491--497

\bibitem {br-tr02}  Briozzo A C and Tarzia D A 2002 An explicit solution for an instantaneous two-phase Stefan problem with nonlinear thermal coefficients {\it IMA Journal of Applied Mathematics} \textbf{67} 249--261

 \bibitem {voller04} Voller V R, Swenson J B and Paola C 2004 An
 analytical solution for a Stefan problem with variable latent heat
 {\it Int. J. Heat Mass Transfer} \textbf{47} 5387--5390

 \bibitem {bl-1974} Bluman G 1974 Application of the general
 similarity solution of the heat equation to boundary-value problems {\it Quart. Appl. Math.}
 \textbf{31} 403--415


\end{thebibliography}
\end{document}